\documentclass[prb,floatfix,preprint,superscriptaddress,showpacs,amsfonts,amssymb,amsmath]{revtex4}
\usepackage{graphicx}
\usepackage{dcolumn}
\usepackage{bm}
\usepackage{subfigure}

\begin{document}
\preprint{Preprint submitted to Phys. Rev. B (05/2006)}
\title{Magnetic anisotropy of Co$_x$Pt$_{1-x}$ clusters embedded in matrix: Influences of the cluster chemical composition and the matrix nature}

\author{S. Rohart      }   \email{srohart@lpmcn.univ-lyon1.fr}
                           \affiliation{Laboratoire de Physique de la Mati\`ere Condens\'ee et Nanostructures, Universit\'e Lyon 1;
                           CNRS, UMR 5586, Domaine Scientifique de la Doua, F-69622 Villeurbanne Cedex; France}
\author{C. Raufast     }   \affiliation{Laboratoire de Physique de la Mati\`ere Condens\'ee et Nanostructures, Universit\'e Lyon 1;
                           CNRS, UMR 5586, Domaine Scientifique de la Doua, F-69622 Villeurbanne Cedex; France}
\author{L. Favre       }   \altaffiliation[Current address: ]{L2MP, Universit\'e Aix-Marseille III; CNRS UMR 6137, F-13397 Marseille Cedex 20; France}
                           \affiliation{Laboratoire de Physique de la Mati\`ere Condens\'ee et Nanostructures, Universit\'e Lyon 1;
                           CNRS, UMR 5586, Domaine Scientifique de la Doua, F-69622 Villeurbanne Cedex; France}
\author{E. Bernstein   }   \affiliation{Laboratoire de Physique de la Mati\`ere Condens\'ee et Nanostructures, Universit\'e Lyon 1;
                           CNRS, UMR 5586, Domaine Scientifique de la Doua, F-69622 Villeurbanne Cedex; France}
\author{E. Bonet       }   \affiliation{Laboratoire Louis N\'eel; CNRS, UPR 5051, BP 166, F-38042 Grenoble Cedex 9; France }
\author{V. Dupuis      }   \affiliation{Laboratoire de Physique de la Mati\`ere Condens\'ee et Nanostructures, Universit\'e Lyon 1;
                           CNRS, UMR 5586, Domaine Scientifique de la Doua, F-69622 Villeurbanne Cedex; France}

\date{\today}

\begin{abstract}
We report on the magnetic properties of Co$_x$Pt$_{1-x}$ clusters
embedded in various matrices. Using a careful analysis of
magnetization curves and ZFC susceptibility measurements, we
determine the clusters magnetic anisotropy energy (MAE) and
separate the surface and volume contributions. By comparing
different chemical compositions, we show that a small amount of Pt
(15~\%) induces an important increase in the volume anisotropy
with respect to pure Co clusters, even in chemically disordered
fcc clusters. Comparing the measurements of  clusters embedded in
Nb and MgO matrices, we show that the oxide matrix induces an
important increase of the surface MAE attributed to the formation
of an antiferromagnetic CoO shell around the clusters.
\end{abstract}


\maketitle

\section{Introduction}

Nanometer-sized magnetic particles have been attracting an
increasing interest over the last decades as their properties
considerably differ from those of bulk materials due to the non
negligible fraction of atoms located at the surfaces or the
interfaces. Enhancement of the orbital magnetic moment as well as
the magnetic anisotropy energy (MAE) at the less coordinated atoms
were observed on many system, from single
atoms,\cite{gambardella2003}
free,\cite{bucher1991,billas1994,xie2003}
supported\cite{durr1999,rusponi2003,weiss2005} and embedded
clusters.\cite{jamet2001,bansmann2004} From a technological point
of view, such nanostructures are potential candidates to increase
the storage media density.\cite{plumer2001} However, applications
are limited by superparamagnetism: due to their reduced sizes, the
nanoparticles MAE is not sufficient to stabilize the magnetization
direction, which fluctuates due to thermal activation. Then, the
need for nanomagnets with higher thermal stability drives the need
for high MAE nanomaterials.

In very small nanostructures, where surface to volume atom number
ratio is not negligible, two contributions to the MAE are found
originating from volume and surface or
interface.\cite{jamet2001,jamet2004} The volume anisotropy is
mainly due to the clusters crystallographic structure, via the
magnetocrystalline anisotropy (MCA). As atomic stacking with a
high symmetry (like cubic staking) does not favor a high MCA,
intense efforts are devoted to the production of mixed clusters
made of a magnetic material (Co, Fe, Ni) with a $4d$ or $5d$
transition metal (Pd,
Pt).\cite{pastor1995,szunyogh1999,bansmann2004} When ordered in
the $L1_0$ tetragonal phase, such mixed bi-metallic systems show
very strong MCA.\cite{sun2000,chen2002,petit2004} The surface
anisotropy has two origins. On the one hand, as the lower
coordinated atoms at the surface are in a less symmetric
environment, they present an enhanced MAE as compared to the
bulk.\cite{jamet2001,jamet2004,gambardella2003} On the other hand,
the contact with a non ferromagnetic matrix induces an interfacial
anisotropy, whose origin depends on the matrix nature. In the case
of metallic matrices, the interfacial anisotropy is due to the
spin-orbit coupling and hybridization between cluster and matrix
atom orbitals, as shown in Co/Pt multilayers\cite{nakajima1998}
and  Co clusters embedded in Pt.\cite{jamet2001b} In the case of
antiferromagnetic matrices, the interfacial anisotropy is due to
the exchange bias phenomenon\cite{meiklejohn1956} as shown for Co
cluster embedded in CoO.\cite{skumryev2003}

In this study, we present magnetic measurements on mixed
Co$_x$Pt$_{1-x}$ clusters (with $x$ ranging from 0 to 1) embedded
in two different matrices: a metallic one (Nb) and an oxide one
(MgO). With a careful analysis of hysteresis loops and Zero Field
Cooled (ZFC) susceptibility measurements, we separate the volume
and surface magnetic anisotropy energies. We then show that, even
in non chemically ordered fcc clusters, the addition of platinum
increases the volume anisotropy with respect to pure Co. We also
show that a high surface anisotropy is found when clusters are
embedded in the oxide matrix, due to the formation of a CoO shell
around the clusters.

\section{Sample elaboration and characterization}

The samples are elaborated using the co-deposition of preformed
clusters in the gas phase and of an atomic flux for the
matrix.\cite{pellarin1994,perez1997} The clusters are produced by
the condensation of a plasma obtained by laser vaporization on a
metallic rod.\cite{milani1990} We use a Nd:YAG laser ($\lambda$ =
532~nm, pulse duration of a few nanoseconds, frequency up to
30~Hz) to vaporize a mixed Co$_x$Pt$_{1-x}$ rod. A continuous He
flux (about 20~mbar) is injected in the vaporization chamber to
cool down the plasma. Clusters nucleate and are stabilized during
a supersonic expansion at the exit nozzle of the source. It
produces a cluster beam of about $10^{-3}$~cluster/nm$^2$/s.
Clusters are deposited in the low energy cluster beam deposition
(LECBD) regime\cite{perez1997} in a UHV chamber (base pressure
$5\times10^{-10}$~mbar) on a Si(001) substrate. Due to their low
kinetic energy, clusters do not fragment upon impact and conserve
their morphology when they are deposited.\cite{perez1997} The
matrix is evaporated using an electron gun evaporator. Nb is
evaporated at 0.2~nm/s and MgO is evaporated at 0.02~nm/s. The
pressure during the co-deposition is below $5\times10^{-8}$~mbar
and falls down rapidly after the process.

\begin{figure}
\hspace{-0.3 cm}
\includegraphics[width=7 cm]{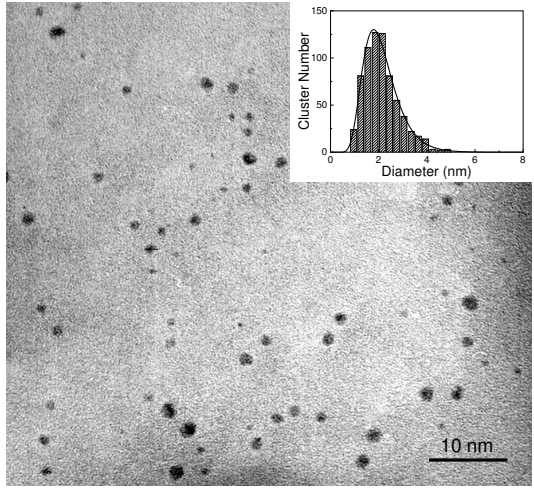}
    \caption{TEM micrograph of the Co$_{85}$Pt$_{15}$ clusters
    deposited on an amorphous carbon grid. Inset: diameter
    distribution. The histogram corresponds to the TEM measurements
    and the continuous line corresponds to a Log-normal fit.} \label{fig_TEM}
\end{figure}

We have produced four types of clusters, using four target rods
with different compositions (cf. table~\ref{tab1}). Using energy
dispersive x-rays (EDX) and Rutherford back scattering (RBS)
spectroscopy measurements, we have characterized the clusters
composition. Our production technique produces mixed clusters with
roughly the same composition as the
target.\cite{rousset1995,rousset2000,gaudry2001,favre2004} As
shown in table~\ref{tab1}, the general tendency is to produce
Co-enriched clusters. This phenomenon may be due to a predominance
of Co atoms evaporation upon laser impact or a re-evaporation of
Pt atoms during the expansion. Then the studied clusters have the
following compositions: Co, Co$_{85}$Pt$_{15}$,
Co$_{58}$Pt$_{42}$, Co$_{30}$Pt$_{70}$ (the indices indicate the
element proportion en percent). The morphology, crystallography
and size distribution of the clusters have been determined by
transmission electron microscopy (TEM) experiments, performed on
clusters deposited on microscopy grids coated by an amorphous
carbon film. A typical image and diameter distribution is shown in
figure~\ref{fig_TEM} for the Co$_{85}$Pt$_{15}$ clusters. In each
case, the diameters follow a Log-normal distribution. The mean
diameter is slightly higher for pure Co clusters (3.2~nm) than for
the mixed clusters (about 2~nm - see table~\ref{tab1}). The
crystallography was determined by the electron diffraction
patterns and high resolution TEM images (HRTEM). In the case of
the pure Co clusters, the HRTEM images show a \{111\}
interreticular distance of about (2.04$\pm$0.02)~\AA, indicating a
fcc crystalline structure. In the case of the mixed clusters, the
diffraction patterns show a \{111\} interreticular distance, which
lies between the Co fcc (2.05~\AA) and Pt fcc (2.27~\AA) \{111\}
interreticular distances, typical of disordered alloys. The fcc
structure is further confirmed by the HRTEM images, which show
that clusters have a faceted truncated octahedron
morphology,\cite{jamet2001,favre2006} corresponding to the typical
equilibrium shape of fcc nanocystallites.\cite{vanhardeveld1969}

\begin{table}[h]
    \begin{ruledtabular}
    \caption{\label{tab1}Isolated clusters morphological characteristics depending on the
    target rod. $D_m$ corresponds to the median diameter and $\sigma$ to
    the dispersion. The composition is determined using EDX spectroscopy. The values are in good agreement (less than 3~\% error) with RBS measurement.} 
    \begin{tabular}{llll}
        Target rod    &  Cluster  & $D_m$ & $\sigma$  \\
        \hline
        Co            & Co                    & 3.2 nm   & 0.25 \\
        Co$_3$Pt      & Co$_{85}$Pt$_{15}$    & 2.0 nm   & 0.35 \\
        CoPt          & Co$_{58}$Pt$_{42}$    & 1.95 nm  & 0.35 \\
        CoPt$_3$      & Co$_{30}$Pt$_{70}$    & 1.9 nm   & 0.3
    \end{tabular}
    \end{ruledtabular}
\end{table}

In a second step, we have considered the clusters embedded in the
matrices. We have used several characterization x-rays based
techniques.\cite{jamet2000,favre2006} These measurements have
shown that, from the crystallographic point of view, the interface
is not sharp. From extended x-rays absorption fine structures
(EXAFS) measurement and simulations on Co clusters embedded in a
Nb matrix,\cite{jamet2000} we show that an intermixing occurs on
the two first layers leading to a CoNb shell, reducing the pure Co
cluster diameter. From grazing incidence x-rays small and wide
angle scattering (GISAXS and GIWAXS) on Co$_{58}$Pt$_{42}$
clusters embedded in a MgO matrix,\cite{favre2006} the occurrence
of the set of Bragg peaks confirms the presence of only one
alloyed phase with a \{111\} interreticular distance of
2.21~\AA~but with a mean CoPt nanocrystallite diameter of 1.3~nm.
This result is due to the oxidation of the cluster surface, which
has been shown by x-ray absorption spectroscopy at the $L_{2,3}$
Co edges. Then, the clusters have a CoPt core and a Co(Pt)O shell.
From such a diffuse cluster surface, one could expect
inhomogeneous magnetic properties. Surprisingly, it was already
shown from microSQUID measurements performed on single Co clusters
embedded in a Nb matrix that even if the CoNb shell is not
magnetic, the Co core conserves the magnetic properties of a well
defined faceted cluster.\cite{jamet2000,jamet2001,jamet2004} In
the following we show that the magnetic properties of our clusters
are also quite homogeneous.

\section{Magnetic measurements and quantitative analysis}

The magnetic measurements have been performed using a
superconducting quantum interference device (SQUID) at various
temperatures (except for the Co:Nb sample, which was measured
using a vibrating sample magnetometer.\cite{jamet2000}) In order
to extract the single isolated cluster magnetic behavior and avoid
any cluster magnetic interactions, we have studied low
concentrated samples. For the clusters embedded in the Nb matrix
the concentration was about 0.1~\% in volume and for the clusters
embedded in the MgO matrix, the concentration was about 5~\% in
volume. At these concentrations the mean distance between clusters
is about 15~nm in Nb and 5~nm in MgO. This is sufficiently large
to discard Ruderman-Kittel-Kasuya-Yasuda interactions in metallic
matrices, which vanish above a few
nanometers.\cite{grolier1993,bruno1991} Dipolar interactions can
also be neglected as the typical interaction
temperature\cite{allia2001} is much lower than our measurement
temperatures (respectively 0.2~K in Nb and 4~K in MgO). We have
measured the four cluster types in both matrices. In the
following, samples are referred as C:M with C=Co,
Co$_{85}$Pt$_{15}$, Co$_{58}$Pt$_{42}$ or Co$_{30}$Pt$_{70}$ the
cluster type and M=Nb or MgO the matrix.

\subsection{Experimental results}

The $M(H)$ curves were recorded at various temperatures, from 2~K
to 300~K, and for $\mu_0H=-5$ to $+5$~T (field sweeping rate
1~mT/s). Except for the Co$_{30}$Pt$_{70}$ clusters, we have
always observed a clear ferromagnetic signal at low temperature.
The weak signal for the Co$_{30}$Pt$_{70}$ clusters may be due to
a too low ordering temperature, as it is well known that the
Co$_x$Pt$_{1-x}$ bulk Curie temperature decreases with increasing
the Pt proportion.\cite{kootte1991} Therefore, we only present in
the following the results for the Co, Co$_{85}$Pt$_{15}$ and
Co$_{58}$Pt$_{42}$ clusters.

\begin{figure}
\hspace{-0.3 cm}
\includegraphics[width=7 cm]{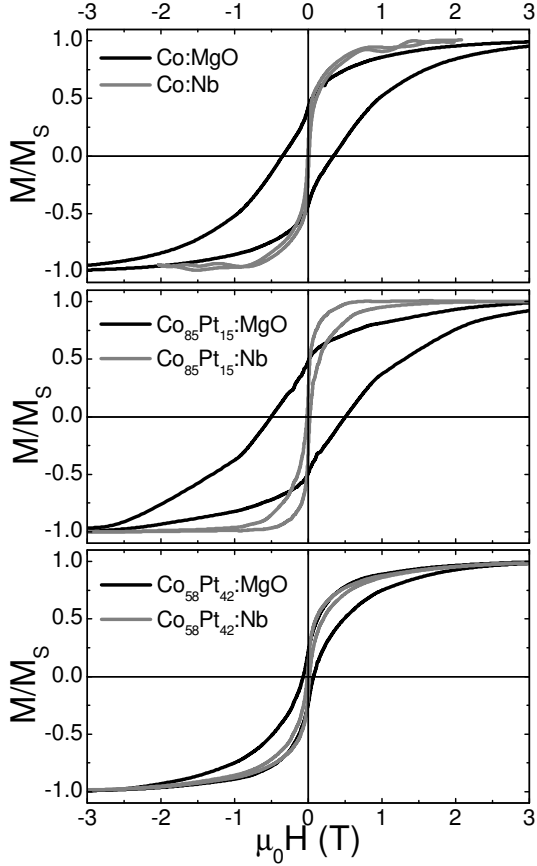}
\caption{Normalized hysteresis loops measured at $T=10$~K on the
samples with Co, Co$_{85}$Pt$_{15}$ and Co$_{58}$Pt$_{42}$
clusters. The dark lines correspond to the measurements in the MgO
matrix and the grey lines correspond to the measurements in the Nb
matrix.} \label{fig_cycle}
\end{figure}

\begin{figure}
\hspace{-0.3 cm}
\includegraphics[width=7 cm]{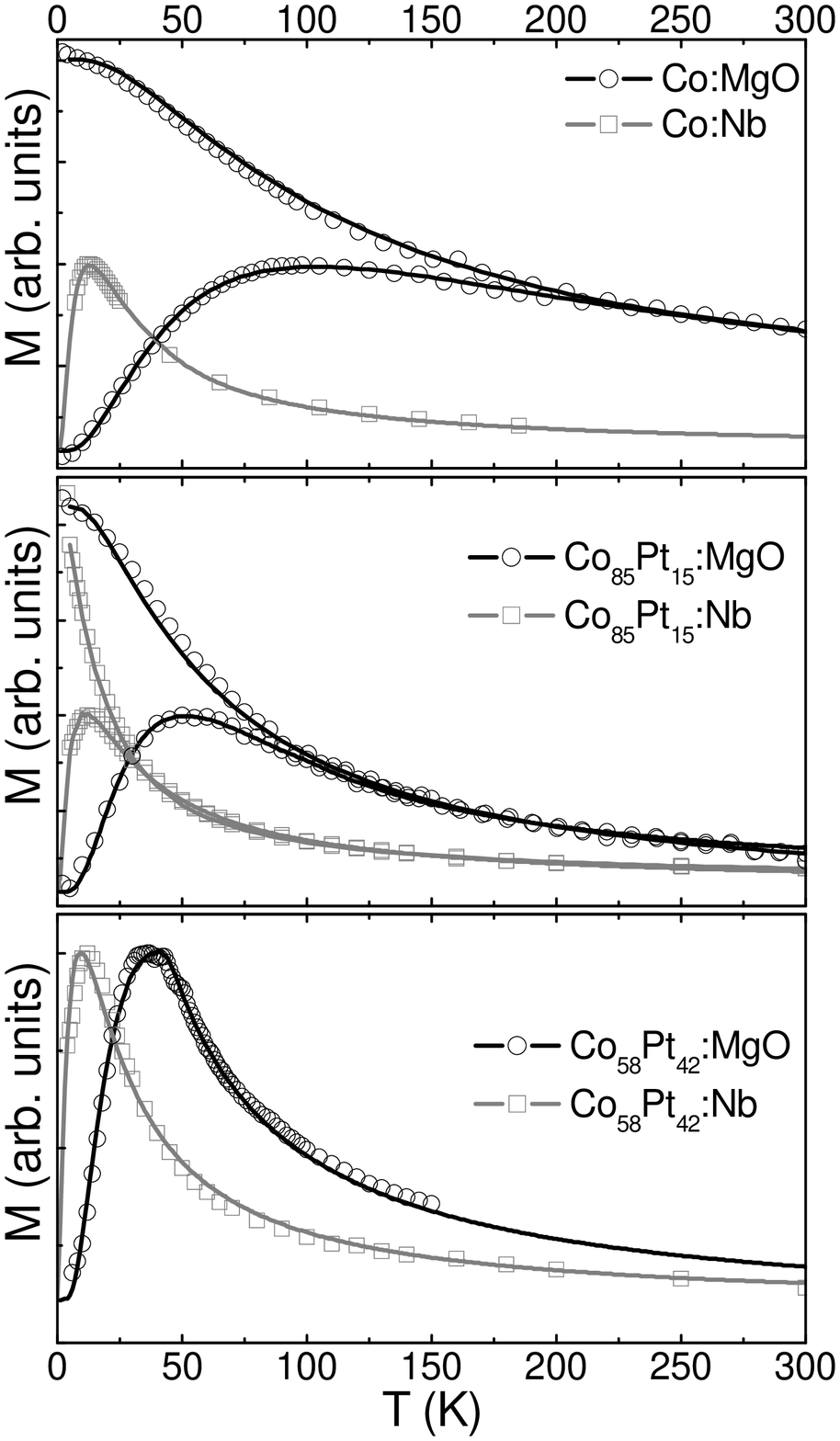}
\caption{ZFC-FC curves measured on the Co, Co$_{85}$Pt$_{15}$ and
Co$_{58}$Pt$_{42}$ clusters embedded in MgO and Nb matrices. The
curves were recorded in a small field of 5~mT (Co:MgO and
Co$_{85}$Pt$_{15}$:MgO) or 10~mT (Co:Nb, Co$_{85}$Pt$_{15}$:Nb,
Co$_{58}$Pt$_{42}$:Nb and Co$_{58}$Pt$_{42}$:MgO). The dots
correspond to the measurements and the continuous lines correspond
to the fits using the model described below.} \label{fig_ZFC}
\end{figure}

At high temperature, the $M(H)$ curves are not hysteretic on any
sample, typical of a superparamagnetic (SPM) behavior as will be
discussed later. On the contrary, the curves taken at low
temperature show open hysteresis loops, typical of a ferromagnetic
or blocked regime (FM). Due to the superconducting transition in
Nb below 8~K, background subtraction becomes hazardous for the
samples with the Nb matrix. Therefore we cannot compare the data
taken at the lowest temperatures and we focus on the curves
measured at 10~K (fig.~\ref{fig_cycle}). The transition between
SPM and FM regimes was characterized using the so-called zero
field cooled-field cooled (ZFC-FC) protocol (fig.~\ref{fig_ZFC}),
recorded in a 5 to 10~mT field, with a temperature sweeping rate
of few tens of mK/s. For every measurement, the ZFC curve shows a
maximum at $T=T_{\mathrm{max}}$, which is related to the blocking
temperature $T_B$ of the particles. For $T>T_{\mathrm{max}}$, we
observe a decrease of $M$, which is proportional to $T^{-1}$,
typical of SPM.

As a general trend, we observe that the peak temperature
$T_{\mathrm{max}}$ in the ZFC curves for a given type of cluster
is systematically higher for clusters embedded in MgO than for
clusters embedded in Nb. In the same way, the coercive fields
$H_C$ and remanent to saturation magnetization ratios $M_R/M_S$
are also systematically higher for clusters embedded in MgO than
for clusters embedded in Nb. This shows that the MgO matrix
induces higher magnetic anisotropy energy than the Nb matrix.
However, the origin of this result is difficult to discuss as the
clusters do not have the same magnetic size in both matrices due
to the intermixing at the interface. In the following, we present
our analysis of the measurements and show a model to fit the
ZFC-FC curves. This analysis enables us to deduce quantitatively
the surface and volume anisotropies, which are independent of the
cluster magnetic volume.

    \subsection{Magnetic size}

\begin{figure}
\hspace{-0.3 cm}
\includegraphics[width=7 cm]{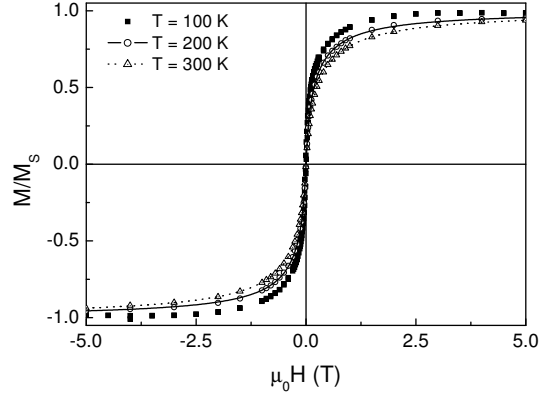}
\caption{Superparamagnetic $M(H)$ curves measured on the Co:MgO
sample at 100~K, 200~K and 300~K. The two measures at the highest
temperatures are fitted using the convolution of a Langevin
function with Log-normal distribution of the magnetic diameter.
The curve recorded at 100~K cannot be fitted with the same
parameters due to a non negligible MAE.} \label{fig_langevin}
\end{figure}

The first point of the analysis is to determine precisely the
clusters magnetic volume and its corresponding distribution in the
samples. For that purpose, we focus on the superparamagnetic
$M(H)$ curves measured at high temperatures ($T\gg
T_{\mathrm{max}}$). At these temperatures, the thermal energy is
high compared to the MAE, which can be neglected.\cite{fruch2002}
The magnetic behavior can be described by the classical Langevin
function. In order to fit the measurements, we take into account
the magnetic volume distribution. Then, the measurements are
described by the convolution of a Langevin function with the
magnetic volume distribution as explained in
Ref.~\onlinecite{jamet2000}. Following the TEM results, we assume
a Log-normal distribution for the cluster magnetic diameter. As
magnetization enhancement only occurs for clusters containing less
than about 500~atoms (about 2~nm in diameter),\cite{billas1994} we
use bulk magnetization values. In pure Co clusters, we take
$m_{Co}=1.7$~$\mu_B$.\cite{jamet2000} In Co$_x$Pt$_{1-x}$ clusters
we take $m_{Co}=1.9$~$\mu_B$ and
$m_{Pt}=0.45$~$\mu_B$.\cite{grange2000} For each sample, we fit at
least two measurements, for $T>2T_{\mathrm{max}}$. An example for
the Co:MgO sample is shown in fig.~\ref{fig_langevin} and the
results are shown in table~\ref{tab2}.

\begin{table*}[t]
    \caption{\label{tab2} Magnetic characteristics of the samples
    deduced from the magnetic measurements and their modeling. The
    magnetic median diameter $D_{m,\mathrm{Mag}}$ and dispersion $\sigma_{\mathrm{Mag}}$
    are obtained from the Langevin fit of the superparamagnetic M(H)
    curves. $T_{\mathrm{max}}$ corresponds to the maximum ZFC susceptibility
    temperature. $H_C$ is the coercive field at 10~K. The remanent
    to saturation magnetization ratio is measured at 10~K. The values in
    parenthesis correspond to the theoretical value calculated from the
    model described below. $K_V$ and $K_S$ are determined by
    fitting the ZFC-FC curves using the same model.}
    \begin{ruledtabular}
    \begin{tabular}{lrrrrrrrr}
        Sample    & \multicolumn{1}{c}{$D_{m,\mathrm{Mag}}$}  & \multicolumn{1}{c}{$\sigma_{\mathrm{Mag}}$} & \multicolumn{1}{c}{$T_{\mathrm{max}}$}
        & \multicolumn{1}{c}{$\mu_0H_C$} & \multicolumn{1}{c}{$M_R/M_S$}
        & \multicolumn{1}{c}{$K_V$}         & \multicolumn{1}{c}{$K_S$}         \\
        \hline
    Co:MgO                 &  2.30 nm     & 0.40           & 100 K &   340 mT   &  0.43 (0.49)  & $(40\pm21)$ kJ/m$^3$  & $(360\pm10)$ $\mu$J/m$^2$  \\
    Co:Nb                  &  2.27 nm     & 0.39           &  12 K &     5 mT   &  0.10 (0.15)  & $(60\pm15)$ kJ/m$^3$  & $( 45\pm5)$ $\mu$J/m$^2$  \\\\
    Co$_{85}$Pt$_{15}$:MgO &  1.45 nm     & 0.38           &  50 K &   500 mT   &  0.49 (0.47)  &$(200\pm19)$ kJ/m$^3$  & $(420\pm11)$ $\mu$J/m$^2$  \\
    Co$_{85}$Pt$_{15}$:Nb  &  1.00 nm     & 0.40           &  13 K &    20 mT   &  0.25 (0.21)  &$(170\pm13)$ kJ/m$^3$  & $(160\pm13)$ $\mu$J/m$^2$  \\\\
    Co$_{58}$Pt$_{42}$:MgO &  1.52 nm     & 0.38           &  39 K &    70 mT   &  0.23 (0.27)  &$(100\pm18)$ kJ/m$^3$  & $(200\pm15)$ $\mu$J/m$^2$  \\
    Co$_{58}$Pt$_{42}$:Nb  &  0.97 nm     & 0.41           &  12 K &    15 mT   &  0.13 (0.13)  &$(110\pm14)$ kJ/m$^3$  & $(170\pm12)$ $\mu$J/m$^2$  \\
    \end{tabular}
    \end{ruledtabular}
\end{table*}

We systematically find a reduced magnetic size as compared to the
TEM measured size (table~\ref{tab1}). This is due to the
cluster-matrix interface, which induces some dead surface layers.
In the case of the clusters embedded in the Nb matrix, the mean
diameter is reduced by about 1~nm, which corresponds to two dead
surface layers. This is coherent with a previous EXAFS study on
the Co:Nb sample,\cite{jamet2000} which has shown the formation of
a non magnetic CoNb alloy for the first two surface layers. X-ray
magnetic circular dichroism (XMCD) measurements on the
Co:Nb\cite{dupuis2003} and Co$_{58}$Pt$_{42}$:Nb
samples\cite{favre2006} have also shown the same results. The mean
spin moments $\mu_\mathrm{S}$ determined from XMCD sum rules are
found to be lower than the bulk spin moment. The ratio
(respectively $\mu_\mathrm{S}/\mu_{\mathrm{S,bulk}}=0.33$ for
Co:Nb and 0.12 for Co$_{58}$Pt$_{42}$:Nb) is close to the ratio of
the magnetic volume to the cluster volume (respectively 0.37 for
Co:Nb and 0.16 for Co$_{58}$Pt$_{42}$:Nb), which corroborates our
magnetic volume determination. In the case of the clusters
embedded in the MgO matrix, the mean diameter is reduced by about
0.5~nm for Co$_{85}$Pt$_{15}$:MgO and Co$_{58}$Pt$_{42}$:MgO and
1~nm for Co:MgO, which corresponds respectively to one and two
dead surface layers. In the case of Co$_{58}$Pt$_{42}$:MgO, this
is coherent with the x-ray diffraction measurements and with the
XMCD.\cite{favre2006} In this last case, the mean spin moment to
bulk spin moment ratio was found to be 0.34,  close to the
magnetic volume to cluster volume ratio (0.41). In this case, the
dead layers are due to the formation of a Co(Pt)O
antiferromagnetic shell around the cluster magnetic core, observed
in the Co L$_3$ multiplet peak in the x-ray absorption peak.

    \subsection{Surface and volume anisotropies }

We now determine the magnetic anisotropy energies in the samples.
For this purpose we focus on the ZFC-FC measurements. In many
studies, the MAE is evaluated by considering that
$T_{\mathrm{max}}$ is the clusters blocking temperature. Whereas
this is true for a monodisperse cluster assembly, such an
assumption leads to an overestimation of the MAE for a distributed
assembly. In the following we expose a model which describes the
ZFC-FC curves for a monodisperse cluster assembly. Then we fit the
measurements, using a convolution of the monodisperse assembly
model with the magnetic volume distribution determined previously.

We consider that our clusters are single domain magnets with a
uniaxial magnetic anisotropy, a quite common assumption for such
nanostructures.\cite{wernsdorfer1997,jamet2001} As no crystalline
direction or orientation is favored during the LECBD and because
the matrix is polycrystalline, the cluster anisotropy axes are
randomly oriented. The cluster magnetic energy is described by the
Stoner-Wohlfarth Hamiltonian.\cite{stoner1948} They can be
magnetized in two directions (named 1 and 2 in the following),
which depend on the anisotropy axis orientation and external
field. The transition from SPM to FM regimes is due to kinetic and
thermal effects. Therefore, the relaxation of the magnetic moments
is described using the rate
equation\cite{chantrell1994,street1994,andersson1997}
\begin{equation}
    \frac{dn_1}{dt}=-\nu_{12}n_1+\nu_{21}(1-n_1),
\end{equation}

\noindent where $n_1$ is the proportion of clusters, whose
magnetization points in the direction 1 and $\nu_{12}$ (resp.
$\nu_{21}$) is the switching rate from direction 1 to 2 (resp.
from 2 to 1). The switching rates are expressed as
$\nu_i=\nu_0\exp(-\Delta E_i/k_BT)$ (with $i=12$ or 21) where
$\Delta E_i$ is the energy barrier and $\nu_0$ the attempt
frequency (about $10^{10}$~Hz).\cite{wernsdorfer1997} These energy
barriers depend on the external magnetic field $H$ and its
orientation with respect to the cluster anisotropy axis.
Unfortunately, no analytical expression can be found for the
energy barriers if the external magnetic field $H$ is not aligned
with the anisotropy axis. However, in the case of a low field
($\mu_0H\ll K/\mu$, with $K$ the cluster MAE and $\mu$ the cluster
magnetic moment), it can be estimated as\cite{andersson1997}
\begin{subequations}
\begin{eqnarray}
 \Delta E_{12} & = & K\left[1-2\frac{H}{H_K}(\sin\psi-\cos\psi)\right]\\
 \Delta E_{21} & = & K\left[1-2\frac{H}{H_K}(\sin\psi+\cos\psi)\right],
\end{eqnarray}
\end{subequations}

\noindent where $\mu_0 H_K=2K/\mu$ is the anisotropy field and
$\psi$ is the angle between the magnetic field and the anisotropy
axis. For an assembly of clusters with random orientation of the
easy axis, the magnetization along the field direction $M$ is the
average of all the possible orientations. If we further assume
$\mu_0 H \ll k_B T / \mu$ and $H$ constant, then the magnetization
$M$ is the solution of
\begin{equation}\label{master}
    \tau(T)\frac{dM}{dt}+M=\frac{\mu_0\mu^2H\eta}{3K}\left(1+\frac{K}{k_BT}\right),
\end{equation}

\noindent with $\eta$ the cluster density and
$\tau(T)=\frac{1}{2\nu_0}\exp(K/k_BT)$, the relaxation time. In
the ZFC-FC measurement protocol, the sample is previously
thermally demagnetized. Then, the initial condition is
$M_0=\mu_0\mu^2H\eta/3K$, which corresponds to the magnetic
susceptibility due to the displacement of the energy minima by the
external magnetic field.\cite{andersson1997}

In order to find the ZFC magnetization $M_\mathrm{ZFC}(T)$, we
hold $T$ constant and integrate eq.~(\ref{master}) for a
temperature $T'$ swept from 0 to $T$. When $T'$ is close to $T \ll
K/k_B$, the relaxation time $\tau(T') \simeq \tau(T)
\exp[(T-T')K/k_BT^2]$ varies on a temperature scale $k_BT'^2/K \ll
T'$ and we can therefore keep the constant T in the right hand of
eq.~(\ref{master}). Given a constant temperature sweeping rate
$dT/dt$, we have then the solution
\begin{equation}\label{ZFC}
M_{\mathrm{ZFC}}(T) = \frac{\mu_0\mu^2H\eta}{3K}\left[
    1 + \frac{K}{k_BT} (1 - e^{-\delta t/\tau(T)})
\right],
\end{equation}
where $\delta t = k_BT^2(KdT/dt)^{-1}$. This is also the
magnetization we would get by first stepping the temperature from
$0$ to $T$ and then letting $M$ relax for a characteristic time
$\delta t$ at constant $T$. We should notice that when $T \ll
K/k_B$ is not true, the solution (\ref{ZFC}) still holds, since in
this regime $M_\mathrm{ZFC}$ has always its equilibrium value.

In order to take into account the magnetic size distribution in
the fitting procedure, we need to know the relation between size
and MAE, which is not trivial as surface and interface effects are
known to play an important role at this size
range.\cite{jamet2001,rusponi2003} We then write the MAE as a
combination of volume and surface anisotropies:
$K=\frac{\pi}{6}D^3K_V+\pi D^2K_S$, where $K_V$ and $K_S$ are
respectively the volume and surface anisotropies. In practice, we
use eq.~(\ref{ZFC}) to fit the experimental ZFC curves and
determine $K_V$ and $K_S$. Then we use a numerical integration of
eq.~(\ref{master}), with the previously determined parameters to
simulate the FC curves. The results of the modelling are shown in
figure~\ref{fig_ZFC} and the fit parameters are reported for each
sample in table~\ref{tab2}.

Equation~(\ref{master}) does not allow performing a calculation of
the whole hysteresis loops, as it is only valid at low magnetic
fields. However, it allows the determination of the remanence to
saturation magnetization ratio. At $T=0$~K, this ratio should be
$\frac{1}{2}$ for an assembly of cluster with randomly oriented
anisotropy axis.\cite{jamet2000} Considering that the
magnetization relaxation is non negligible for $\mu_0 \mu H<k_BT$,
the measured value corresponds to the magnetization reached after
the typical time $\delta t'$ in zero field:
\begin{equation}\label{MR}
    \frac{M_R(T)}{M_S}=\frac{M_R(T=0\mathrm{~K})}{M_S}e^{-\delta t'/\tau(T)}=\frac{1}{2}e^{-\delta t'/\tau(T)},
\end{equation}

\noindent where $\delta t'=k_BT(\mu_0\mu dH/dt)^{-1}$. Taking into
account the volume distribution, and using the $K_S$ and $K_V$
values obtained previously, we have calculated
$M_R/M_S(T=10\mathrm{~K})$ for each sample. The results shown in
table~\ref{tab2} are in quite good agreement with the experimental
values.

\section{Discussion}

The volume anisotropies found for a given cluster composition in
both matrices are nearly found equal. This result justifies our
separation of volume and surface anisotropies: the volume
anisotropy is only linked to the type of cluster whereas the
surface anisotropy is related to the cluster surface and the
matrix nature. We find that the mixed clusters have significantly
higher volume anisotropy as compared to the pure Co clusters.
Then, although no chemical ordering was found in the mixed
clusters, the Pt atoms still favors a MAE enhancement. This result
could be attributed to a finite size effect: due to the small
cluster size, the atoms distribution cannot be totaly homogeneous
in the cluster and a small chemical anisotropy can be observed,
leading to an enhanced anisotropy, via the hybridization between
Co $3d$ and Pt $5d$ orbitals. $K_V$ is found to be maximum for the
Co$_{85}$Pt$_{15}$ clusters (three times the Co volume
anisotropy). Note that for the chemical ordered phases a higher
volume anisotropy is expected for the L1$_0$ CoPt phase than for
the L1$_2$ Co$_3$Pt one. As we observe the opposite tendency, the
origin of the higher $K_V$ in the Co$_{85}$Pt$_{15}$ could be due
to a small chemical ordering in the cluster core. This hypothesis
cannot be proved up to now but we are currently performing high
resolution EDX and TEM measurements, and simulations to obtain a
more precise crystallographic characterization.

Concerning the surface anisotropies, the variation according to
the cluster composition is non-trivial. On the one hand, for
clusters embedded in Nb, we observe that $K_S$ increases with the
proportion of Pt. On the other hand, for clusters embedded in MgO,
$K_S$  is maximum for the Co$_{85}$Pt$_{15}$. This complex result
comes from the fact that $K_S$ has two origins. The first one
($K_{S,\mathrm{C}}$) is intrinsic to the cluster (independent of
the matrix) and is due to the low symmetry and atomic coordination
at the cluster surface.\cite{gambardella2003,jamet2004} The second
one ($K_{S,\mathrm{C/M}}$) is due to the cluster/matrix interface
and then depends on both cluster and matrix.
It was already shown that the interface anisotropy for Co clusters
embedded in Nb is negligible.\cite{jamet2001,jamet2004} Assuming
that this fact is still valid for the mixed clusters, it is
possible to estimate the different contributions from our
measurements (see table~\ref{tab3}). From this estimation we find
that the Pt atoms induce an enhancement of the cluster intrinsic
surface anisotropy as compared to pure clusters. This result is
attributed to the presence of Pt atoms at the surface, which
decrease the coordination number of the cobalt surface atoms.

\begin{table}[h]
    \caption{\label{tab3}Intrinsic and interface anisotropy of the
    cluster embedded in Nb and MgO estimated from the measurements
    (table~\ref{tab2}) as $K_S(\mathrm{C:M})=K_{S,\mathrm{C}}+K_{S,\mathrm{C/M}}$. The
    interface anisotropy of cluster embedded in Nb is supposed to
    be zero.}
    \begin{ruledtabular}
    \begin{tabular}{lrrr}
    Cluster        & $K_{S,\mathrm{C}}$& $K_{S,\mathrm{C/Nb}}$ & $K_{S,\mathrm{C/MgO}}    $\\\hline
    Co             &  45 $\mu$J/m$^2$ & 0            & 315 $\mu$J/m$^2$\\
    Co$_{85}$Pt$_{15}$ & 160 $\mu$J/m$^2$ & 0            & 260 $\mu$J/m$^2$\\
    Co$_{58}$Pt$_{42}$ & 170 $\mu$J/m$^2$ & 0            &  30 $\mu$J/m$^2$\\
    \end{tabular}
    \end{ruledtabular}
\end{table}

We now discuss the origin of the interface anisotropy of the
clusters embedded in MgO. We note that the very high surface
anisotropy found for the pure Co clusters embedded in MgO matrix
is very close to the one found in a Pt matrix
($K_S($Co:Pt$)=300$~$\mu$J/m$^2$).\cite{jamet2001b} However, the
origin is completely different in the two cases. In the case of
the Pt matrix, the $K_S$ enhancement is due to the hybridization
between Co and Pt atoms at the interface, a well known effect in
the Co/Pt multilayers.\cite{hillebrands1993,nakajima1998} Such an
effect is not possible in the case of the MgO matrix. The origin
is rather due to the formation of an oxide shell around the
clusters, as it was observed in the x-ray absorption
spectra.\cite{favre2006} It is well known that the interface
between an antiferromagnetic (AFM) and a ferromagnetic layer leads
to the so-called exchange anisotropy and to the exchange bias
phenomenon.\cite{meiklejohn1956} Concerning nanoparticles, this
effect was shown to induce an important increase of the
anisotropy.\cite{skumryev2003,morel2004} In order to evidence the
exchange bias phenomena, we have measured the low temperature
hysteresis loops after a field cooled under a high magnetic field.
After a 3~T field cooled, the Co$_{58}$Pt$_{42}$:MgO sample has
shown a small $12\pm5$~mT exchange bias field at 6~K. After a 6~T
field cooled, the Co:MgO sample has shown no exchange bias at 2~K.
This unclear exchange bias observation may be related to the small
thickness of the CoO shell. Indeed, in such a case, the AFM shell
MAE is too low as compared to the coupling energy between the FM
core and AFM shell. Then the magnetic moments in the shell rotate
coherently with the FM core\cite{dobrynin2005} and the energy to
overcome in order to reverse the cluster magnetization corresponds
to the sum of the core and shell MAE. Concerning the cluster/MgO
interface anisotropy variation with the cluster chemical
composition, we remark that it decreases when the Pt proportion is
increased and we find that this contribution is weak for the
Co$_{58}$Pt$_{42}$ clusters. We attribute this result to the
presence of Pt atoms at the cluster surface, which decrease the
quality of the antiferromagnetic ordering in the oxide shell, as
it was already observed in ferromagnetic/antiferromagnetic
multilayers.\cite{hong2006}

\section{Conclusion}

In conclusion, we have studied the magnetic anisotropy of well
defined Co$_x$Pt$_{1-x}$ nanoclusters embedded in Nb and MgO
matrices. Using a careful analysis of $M(H)$ curves and ZFC-FC
susceptibility measurements, we have determined the magnetic
anisotropy of the clusters and separated the surface and volume
contributions. By comparing samples of different compositions, we
have shown that a small amount of Pt (15~\%) in the clusters is
responsible for a strong increase of the volume anisotropy, even
if the clusters are fcc chemically disordered. Once more, we have
shown that the MgO matrix induces an enhanced surface anisotropy
due to the partial oxidation of the surface of the clusters.

\begin{acknowledgments}
The authors are indebted to O. Boisron, G. Guiraud and C. Clavier
for their continuous technical assistance and developments during
LECBD experiments and to E. Eyraud for his technical assistance
during the SQUID measurements. We acknowledge support from the
European Community (STREP SFINx no. NMP2-CT-2003-505587).
\end{acknowledgments}


\begin{thebibliography}{46}
\expandafter\ifx\csname
natexlab\endcsname\relax\def\natexlab#1{#1}\fi
\expandafter\ifx\csname bibnamefont\endcsname\relax
  \def\bibnamefont#1{#1}\fi
\expandafter\ifx\csname bibfnamefont\endcsname\relax
  \def\bibfnamefont#1{#1}\fi
\expandafter\ifx\csname citenamefont\endcsname\relax
  \def\citenamefont#1{#1}\fi
\expandafter\ifx\csname url\endcsname\relax
  \def\url#1{\texttt{#1}}\fi
\expandafter\ifx\csname
urlprefix\endcsname\relax\def\urlprefix{URL }\fi
\providecommand{\bibinfo}[2]{#2}
\providecommand{\eprint}[2][]{\url{#2}}

\bibitem[{\citenamefont{Gambardella et~al.}(2003)\citenamefont{Gambardella,
  Rusponi, Veronese, Dehsi, Grazioli, Dallmeyer, Cabria, Zeller, Dederichs,
  Kern et~al.}}]{gambardella2003}
\bibinfo{author}{\bibfnamefont{P.}~\bibnamefont{Gambardella}},
  \bibinfo{author}{\bibfnamefont{S.}~\bibnamefont{Rusponi}},
  \bibinfo{author}{\bibfnamefont{M.}~\bibnamefont{Veronese}},
  \bibinfo{author}{\bibfnamefont{S.~S.} \bibnamefont{Dehsi}},
  \bibinfo{author}{\bibfnamefont{C.}~\bibnamefont{Grazioli}},
  \bibinfo{author}{\bibfnamefont{A.}~\bibnamefont{Dallmeyer}},
  \bibinfo{author}{\bibfnamefont{I.}~\bibnamefont{Cabria}},
  \bibinfo{author}{\bibfnamefont{R.}~\bibnamefont{Zeller}},
  \bibinfo{author}{\bibfnamefont{P.~H.} \bibnamefont{Dederichs}},
  \bibinfo{author}{\bibfnamefont{K.}~\bibnamefont{Kern}}, \bibnamefont{et~al.},
  \bibinfo{journal}{Science} \textbf{\bibinfo{volume}{300}},
  \bibinfo{pages}{1130} (\bibinfo{year}{2003}).

\bibitem[{\citenamefont{Bucher et~al.}(1991)\citenamefont{Bucher, Douglass, and
  Bloomfield}}]{bucher1991}
\bibinfo{author}{\bibfnamefont{J.~P.} \bibnamefont{Bucher}},
  \bibinfo{author}{\bibfnamefont{D.~C.} \bibnamefont{Douglass}},
  \bibnamefont{and} \bibinfo{author}{\bibfnamefont{L.~A.}
  \bibnamefont{Bloomfield}}, \bibinfo{journal}{Phys. Rev. Lett.}
  \textbf{\bibinfo{volume}{66}}, \bibinfo{pages}{3052} (\bibinfo{year}{1991}).

\bibitem[{\citenamefont{Billas et~al.}(1994)\citenamefont{Billas, Ch\^{a}telin,
  and de~Heer}}]{billas1994}
\bibinfo{author}{\bibfnamefont{I.~M.} \bibnamefont{Billas}},
  \bibinfo{author}{\bibfnamefont{A.}~\bibnamefont{Ch\^{a}telin}},
  \bibnamefont{and} \bibinfo{author}{\bibfnamefont{W.~A.}
  \bibnamefont{de~Heer}}, \bibinfo{journal}{Science}
  \textbf{\bibinfo{volume}{265}}, \bibinfo{pages}{1682} (\bibinfo{year}{1994}).

\bibitem[{\citenamefont{Xie and Blackman}(2003)}]{xie2003}
\bibinfo{author}{\bibfnamefont{Y.}~\bibnamefont{Xie}} \bibnamefont{and}
  \bibinfo{author}{\bibfnamefont{J.~A.} \bibnamefont{Blackman}},
  \bibinfo{journal}{J. Phys.: Condens. Matter} \textbf{\bibinfo{volume}{15}},
  \bibinfo{pages}{L615} (\bibinfo{year}{2003}).

\bibitem[{\citenamefont{D\"{u}rr et~al.}(1999)\citenamefont{D\"{u}rr, Dhesi,
  Dudzik, Knabben, van~der Laan, Goedkoop, and Hillebrecht}}]{durr1999}
\bibinfo{author}{\bibfnamefont{H.~A.} \bibnamefont{D\"{u}rr}},
  \bibinfo{author}{\bibfnamefont{S.~S.} \bibnamefont{Dhesi}},
  \bibinfo{author}{\bibfnamefont{E.}~\bibnamefont{Dudzik}},
  \bibinfo{author}{\bibfnamefont{D.}~\bibnamefont{Knabben}},
  \bibinfo{author}{\bibfnamefont{G.}~\bibnamefont{van~der Laan}},
  \bibinfo{author}{\bibfnamefont{J.~B.} \bibnamefont{Goedkoop}},
  \bibnamefont{and} \bibinfo{author}{\bibfnamefont{F.~U.}
  \bibnamefont{Hillebrecht}}, \bibinfo{journal}{Phys. Rev. B}
  \textbf{\bibinfo{volume}{59}}, \bibinfo{pages}{R701} (\bibinfo{year}{1999}).

\bibitem[{\citenamefont{Rusponi et~al.}(203)\citenamefont{Rusponi, Cren, Weiss,
  Epple, Buluschek, Claude, and Brune}}]{rusponi2003}
\bibinfo{author}{\bibfnamefont{S.}~\bibnamefont{Rusponi}},
  \bibinfo{author}{\bibfnamefont{T.}~\bibnamefont{Cren}},
  \bibinfo{author}{\bibfnamefont{N.}~\bibnamefont{Weiss}},
  \bibinfo{author}{\bibfnamefont{M.}~\bibnamefont{Epple}},
  \bibinfo{author}{\bibfnamefont{P.}~\bibnamefont{Buluschek}},
  \bibinfo{author}{\bibfnamefont{L.}~\bibnamefont{Claude}}, \bibnamefont{and}
  \bibinfo{author}{\bibfnamefont{H.}~\bibnamefont{Brune}},
  \bibinfo{journal}{Nature Mater.} \textbf{\bibinfo{volume}{2}},
  \bibinfo{pages}{546} (\bibinfo{year}{203}).

\bibitem[{\citenamefont{Weiss et~al.}(2005)\citenamefont{Weiss, Cren, Epple,
  Rusponi, Baudot, Rohart, Tejeda, Repain, Rousset, Ohresser
  et~al.}}]{weiss2005}
\bibinfo{author}{\bibfnamefont{N.}~\bibnamefont{Weiss}},
  \bibinfo{author}{\bibfnamefont{T.}~\bibnamefont{Cren}},
  \bibinfo{author}{\bibfnamefont{M.}~\bibnamefont{Epple}},
  \bibinfo{author}{\bibfnamefont{S.}~\bibnamefont{Rusponi}},
  \bibinfo{author}{\bibfnamefont{G.}~\bibnamefont{Baudot}},
  \bibinfo{author}{\bibfnamefont{S.}~\bibnamefont{Rohart}},
  \bibinfo{author}{\bibfnamefont{A.}~\bibnamefont{Tejeda}},
  \bibinfo{author}{\bibfnamefont{V.}~\bibnamefont{Repain}},
  \bibinfo{author}{\bibfnamefont{S.}~\bibnamefont{Rousset}},
  \bibinfo{author}{\bibfnamefont{P.}~\bibnamefont{Ohresser}},
  \bibnamefont{et~al.}, \bibinfo{journal}{Phys. Rev. Lett.}
  \textbf{\bibinfo{volume}{95}}, \bibinfo{pages}{157204}
  (\bibinfo{year}{2005}).

\bibitem[{\citenamefont{Jamet et~al.}(2001{\natexlab{a}})\citenamefont{Jamet,
  Wernsdorfer, Thirion, Mailly, Dupuis, M\'elinon, and P\'erez}}]{jamet2001}
\bibinfo{author}{\bibfnamefont{M.}~\bibnamefont{Jamet}},
  \bibinfo{author}{\bibfnamefont{W.}~\bibnamefont{Wernsdorfer}},
  \bibinfo{author}{\bibfnamefont{C.}~\bibnamefont{Thirion}},
  \bibinfo{author}{\bibfnamefont{D.}~\bibnamefont{Mailly}},
  \bibinfo{author}{\bibfnamefont{V.}~\bibnamefont{Dupuis}},
  \bibinfo{author}{\bibfnamefont{P.}~\bibnamefont{M\'elinon}},
  \bibnamefont{and} \bibinfo{author}{\bibfnamefont{A.}~\bibnamefont{P\'erez}},
  \bibinfo{journal}{Phys. Rev. Lett.} \textbf{\bibinfo{volume}{86}},
  \bibinfo{pages}{4676} (\bibinfo{year}{2001}{\natexlab{a}}).

\bibitem[{\citenamefont{Bansmann et~al.}(2004)\citenamefont{Bansmann, Baker,
  Binns, Blackman, Bucher, Dorantes-D\'avila, Dupuis, Favre, Kechrakos,
  Kleibert et~al.}}]{bansmann2004}
\bibinfo{author}{\bibfnamefont{J.}~\bibnamefont{Bansmann}},
  \bibinfo{author}{\bibfnamefont{S.~H.} \bibnamefont{Baker}},
  \bibinfo{author}{\bibfnamefont{C.}~\bibnamefont{Binns}},
  \bibinfo{author}{\bibfnamefont{J.~A.} \bibnamefont{Blackman}},
  \bibinfo{author}{\bibfnamefont{J.~P.} \bibnamefont{Bucher}},
  \bibinfo{author}{\bibfnamefont{J.}~\bibnamefont{Dorantes-D\'avila}},
  \bibinfo{author}{\bibfnamefont{V.}~\bibnamefont{Dupuis}},
  \bibinfo{author}{\bibfnamefont{L.}~\bibnamefont{Favre}},
  \bibinfo{author}{\bibfnamefont{D.}~\bibnamefont{Kechrakos}},
  \bibinfo{author}{\bibfnamefont{A.}~\bibnamefont{Kleibert}},
  \bibnamefont{et~al.}, \bibinfo{journal}{Surf. Sci. Rep.}
  \textbf{\bibinfo{volume}{56}}, \bibinfo{pages}{189} (\bibinfo{year}{2004}).

\bibitem[{\citenamefont{Plumer et~al.}(2001)\citenamefont{Plumer, van Ek, and
  Weller}}]{plumer2001}
\bibinfo{author}{\bibfnamefont{M.~L.} \bibnamefont{Plumer}},
  \bibinfo{author}{\bibfnamefont{J.}~\bibnamefont{van Ek}}, \bibnamefont{and}
  \bibinfo{author}{\bibfnamefont{D.}~\bibnamefont{Weller}},
  \emph{\bibinfo{title}{The physics of ultra-high-density magnetic recording}}
  (\bibinfo{publisher}{Springer}, \bibinfo{address}{Berlin},
  \bibinfo{year}{2001}).

\bibitem[{\citenamefont{Jamet et~al.}(2004)\citenamefont{Jamet, Wernsdorfer,
  Thirion, Dupuis, M\'{e}linon, P\'{e}rez, and Mailly}}]{jamet2004}
\bibinfo{author}{\bibfnamefont{M.}~\bibnamefont{Jamet}},
  \bibinfo{author}{\bibfnamefont{W.}~\bibnamefont{Wernsdorfer}},
  \bibinfo{author}{\bibfnamefont{C.}~\bibnamefont{Thirion}},
  \bibinfo{author}{\bibfnamefont{V.}~\bibnamefont{Dupuis}},
  \bibinfo{author}{\bibfnamefont{P.}~\bibnamefont{M\'{e}linon}},
  \bibinfo{author}{\bibfnamefont{A.}~\bibnamefont{P\'{e}rez}},
  \bibnamefont{and} \bibinfo{author}{\bibfnamefont{D.}~\bibnamefont{Mailly}},
  \bibinfo{journal}{Phys. Rev. B} \textbf{\bibinfo{volume}{69}},
  \bibinfo{pages}{024401} (\bibinfo{year}{2004}).

\bibitem[{\citenamefont{Pastor et~al.}(1995)\citenamefont{Pastor,
  Dorantes-D\'avila, Pick, and Dreyss\'e}}]{pastor1995}
\bibinfo{author}{\bibfnamefont{G.~M.} \bibnamefont{Pastor}},
  \bibinfo{author}{\bibfnamefont{J.}~\bibnamefont{Dorantes-D\'avila}},
  \bibinfo{author}{\bibfnamefont{S.}~\bibnamefont{Pick}}, \bibnamefont{and}
  \bibinfo{author}{\bibfnamefont{H.}~\bibnamefont{Dreyss\'e}},
  \bibinfo{journal}{Phys. Rev. Lett.} \textbf{\bibinfo{volume}{75}},
  \bibinfo{pages}{326} (\bibinfo{year}{1995}).

\bibitem[{\citenamefont{Szunyogh et~al.}(1999)\citenamefont{Szunyogh,
  Weinberger, and Sommers}}]{szunyogh1999}
\bibinfo{author}{\bibfnamefont{L.}~\bibnamefont{Szunyogh}},
  \bibinfo{author}{\bibfnamefont{P.}~\bibnamefont{Weinberger}},
  \bibnamefont{and} \bibinfo{author}{\bibfnamefont{C.}~\bibnamefont{Sommers}},
  \bibinfo{journal}{Phys. Rev. B} \textbf{\bibinfo{volume}{60}},
  \bibinfo{pages}{11910} (\bibinfo{year}{1999}).

\bibitem[{\citenamefont{Sun et~al.}(2000)\citenamefont{Sun, Murray, Weller,
  Folks, and Moser}}]{sun2000}
\bibinfo{author}{\bibfnamefont{S.}~\bibnamefont{Sun}},
  \bibinfo{author}{\bibfnamefont{C.~B.} \bibnamefont{Murray}},
  \bibinfo{author}{\bibfnamefont{D.}~\bibnamefont{Weller}},
  \bibinfo{author}{\bibfnamefont{L.}~\bibnamefont{Folks}}, \bibnamefont{and}
  \bibinfo{author}{\bibfnamefont{A.}~\bibnamefont{Moser}},
  \bibinfo{journal}{Science} \textbf{\bibinfo{volume}{287}},
  \bibinfo{pages}{1989} (\bibinfo{year}{2000}).

\bibitem[{\citenamefont{Chen and Nikles}(2002)}]{chen2002}
\bibinfo{author}{\bibfnamefont{M.}~\bibnamefont{Chen}} \bibnamefont{and}
  \bibinfo{author}{\bibfnamefont{D.}~\bibnamefont{Nikles}},
  \bibinfo{journal}{J. Appl. Phys.} \textbf{\bibinfo{volume}{91}},
  \bibinfo{pages}{8477} (\bibinfo{year}{2002}).

\bibitem[{\citenamefont{Petit et~al.}(2004)\citenamefont{Petit, Rusponi, and
  Brune}}]{petit2004}
\bibinfo{author}{\bibfnamefont{C.}~\bibnamefont{Petit}},
  \bibinfo{author}{\bibfnamefont{S.}~\bibnamefont{Rusponi}}, \bibnamefont{and}
  \bibinfo{author}{\bibfnamefont{H.}~\bibnamefont{Brune}}, \bibinfo{journal}{J.
  Appl. Phys.} \textbf{\bibinfo{volume}{95}}, \bibinfo{pages}{4251}
  (\bibinfo{year}{2004}).

\bibitem[{\citenamefont{Nakajima et~al.}(1998)\citenamefont{Nakajima, Koide,
  Shidara, Miyauchi, Fukutani, Fujimori, Iio, Katayama, N\'yvlt, and
  Suzuki}}]{nakajima1998}
\bibinfo{author}{\bibfnamefont{N.}~\bibnamefont{Nakajima}},
  \bibinfo{author}{\bibfnamefont{T.}~\bibnamefont{Koide}},
  \bibinfo{author}{\bibfnamefont{T.}~\bibnamefont{Shidara}},
  \bibinfo{author}{\bibfnamefont{H.}~\bibnamefont{Miyauchi}},
  \bibinfo{author}{\bibfnamefont{H.}~\bibnamefont{Fukutani}},
  \bibinfo{author}{\bibfnamefont{A.}~\bibnamefont{Fujimori}},
  \bibinfo{author}{\bibfnamefont{K.}~\bibnamefont{Iio}},
  \bibinfo{author}{\bibfnamefont{T.}~\bibnamefont{Katayama}},
  \bibinfo{author}{\bibfnamefont{M.}~\bibnamefont{N\'yvlt}}, \bibnamefont{and}
  \bibinfo{author}{\bibfnamefont{Y.}~\bibnamefont{Suzuki}},
  \bibinfo{journal}{Phys. Rev. Lett.} \textbf{\bibinfo{volume}{81}},
  \bibinfo{pages}{5229} (\bibinfo{year}{1998}).

\bibitem[{\citenamefont{Jamet et~al.}(2001{\natexlab{b}})\citenamefont{Jamet,
  N\'egrier, Dupuis, Tuaillon-Combes, M\'elinon, P\'erez, Wernsdorfer, Barbara,
  and Baguenard}}]{jamet2001b}
\bibinfo{author}{\bibfnamefont{M.}~\bibnamefont{Jamet}},
  \bibinfo{author}{\bibfnamefont{M.}~\bibnamefont{N\'egrier}},
  \bibinfo{author}{\bibfnamefont{V.}~\bibnamefont{Dupuis}},
  \bibinfo{author}{\bibfnamefont{J.}~\bibnamefont{Tuaillon-Combes}},
  \bibinfo{author}{\bibfnamefont{P.}~\bibnamefont{M\'elinon}},
  \bibinfo{author}{\bibfnamefont{A.}~\bibnamefont{P\'erez}},
  \bibinfo{author}{\bibfnamefont{W.}~\bibnamefont{Wernsdorfer}},
  \bibinfo{author}{\bibfnamefont{B.}~\bibnamefont{Barbara}}, \bibnamefont{and}
  \bibinfo{author}{\bibfnamefont{B.}~\bibnamefont{Baguenard}},
  \bibinfo{journal}{J. Magn. Magn. Matters} \textbf{\bibinfo{volume}{237}},
  \bibinfo{pages}{293} (\bibinfo{year}{2001}{\natexlab{b}}).

\bibitem[{\citenamefont{Meiklejohn and Bean}(1956)}]{meiklejohn1956}
\bibinfo{author}{\bibfnamefont{W.~H.} \bibnamefont{Meiklejohn}}
  \bibnamefont{and} \bibinfo{author}{\bibfnamefont{C.~P.} \bibnamefont{Bean}},
  \bibinfo{journal}{Phys. Rev.} \textbf{\bibinfo{volume}{102}},
  \bibinfo{pages}{1413} (\bibinfo{year}{1956}).

\bibitem[{\citenamefont{Skumryev et~al.}(2003)\citenamefont{Skumryev, Stoyanov,
  Zhang, Hadjipanayis, Givord, and Nogu\'es}}]{skumryev2003}
\bibinfo{author}{\bibfnamefont{V.}~\bibnamefont{Skumryev}},
  \bibinfo{author}{\bibfnamefont{S.}~\bibnamefont{Stoyanov}},
  \bibinfo{author}{\bibfnamefont{Y.}~\bibnamefont{Zhang}},
  \bibinfo{author}{\bibfnamefont{G.}~\bibnamefont{Hadjipanayis}},
  \bibinfo{author}{\bibfnamefont{D.}~\bibnamefont{Givord}}, \bibnamefont{and}
  \bibinfo{author}{\bibfnamefont{J.}~\bibnamefont{Nogu\'es}},
  \bibinfo{journal}{Nature} \textbf{\bibinfo{volume}{423}},
  \bibinfo{pages}{850} (\bibinfo{year}{2003}).

\bibitem[{\citenamefont{Pellarin et~al.}(1994)\citenamefont{Pellarin,
  Cottancin, Lerm\'e, Vialle, Wolf, Broyer, Paillard, Dupuis, P\'erez, P\'erez
  et~al.}}]{pellarin1994}
\bibinfo{author}{\bibfnamefont{M.}~\bibnamefont{Pellarin}},
  \bibinfo{author}{\bibfnamefont{E.}~\bibnamefont{Cottancin}},
  \bibinfo{author}{\bibfnamefont{J.}~\bibnamefont{Lerm\'e}},
  \bibinfo{author}{\bibfnamefont{J.~L.} \bibnamefont{Vialle}},
  \bibinfo{author}{\bibfnamefont{J.~P.} \bibnamefont{Wolf}},
  \bibinfo{author}{\bibfnamefont{M.}~\bibnamefont{Broyer}},
  \bibinfo{author}{\bibfnamefont{V.}~\bibnamefont{Paillard}},
  \bibinfo{author}{\bibfnamefont{V.}~\bibnamefont{Dupuis}},
  \bibinfo{author}{\bibfnamefont{A.}~\bibnamefont{P\'erez}},
  \bibinfo{author}{\bibfnamefont{J.}~\bibnamefont{P\'erez}},
  \bibnamefont{et~al.}, \bibinfo{journal}{Chem. Phys. Lett.}
  \textbf{\bibinfo{volume}{224}}, \bibinfo{pages}{338} (\bibinfo{year}{1994}).

\bibitem[{\citenamefont{P\'erez et~al.}(1997)\citenamefont{P\'erez, M\'elinon,
  Dupuis, Jensen, Pr\'evel, Broyer, Pellarin, Cottancin, Lerm\'e, Vialle
  et~al.}}]{perez1997}
\bibinfo{author}{\bibfnamefont{A.}~\bibnamefont{P\'erez}},
  \bibinfo{author}{\bibfnamefont{P.}~\bibnamefont{M\'elinon}},
  \bibinfo{author}{\bibfnamefont{V.}~\bibnamefont{Dupuis}},
  \bibinfo{author}{\bibfnamefont{P.}~\bibnamefont{Jensen}},
  \bibinfo{author}{\bibfnamefont{B.}~\bibnamefont{Pr\'evel}},
  \bibinfo{author}{\bibfnamefont{M.}~\bibnamefont{Broyer}},
  \bibinfo{author}{\bibfnamefont{M.}~\bibnamefont{Pellarin}},
  \bibinfo{author}{\bibfnamefont{E.}~\bibnamefont{Cottancin}},
  \bibinfo{author}{\bibfnamefont{J.}~\bibnamefont{Lerm\'e}},
  \bibinfo{author}{\bibfnamefont{J.~L.} \bibnamefont{Vialle}},
  \bibnamefont{et~al.}, \bibinfo{journal}{J. Phys. D}
  \textbf{\bibinfo{volume}{30}}, \bibinfo{pages}{1} (\bibinfo{year}{1997}).

\bibitem[{\citenamefont{Milani and de~Heer}(1990)}]{milani1990}
\bibinfo{author}{\bibfnamefont{P.}~\bibnamefont{Milani}} \bibnamefont{and}
  \bibinfo{author}{\bibfnamefont{W.~A.} \bibnamefont{de~Heer}},
  \bibinfo{journal}{Rev. Sci. Instrum.} \textbf{\bibinfo{volume}{61}},
  \bibinfo{pages}{1835} (\bibinfo{year}{1990}).

\bibitem[{\citenamefont{Rousset et~al.}(2000)\citenamefont{Rousset, Aires,
  Sekhar, M\'elinon, Pr\'evel, and Pellarin}}]{rousset1995}
\bibinfo{author}{\bibfnamefont{J.~L.} \bibnamefont{Rousset}},
  \bibinfo{author}{\bibfnamefont{F.~J. C.~S.} \bibnamefont{Aires}},
  \bibinfo{author}{\bibfnamefont{B.~R.} \bibnamefont{Sekhar}},
  \bibinfo{author}{\bibfnamefont{P.}~\bibnamefont{M\'elinon}},
  \bibinfo{author}{\bibfnamefont{P.}~\bibnamefont{Pr\'evel}}, \bibnamefont{and}
  \bibinfo{author}{\bibfnamefont{M.}~\bibnamefont{Pellarin}},
  \bibinfo{journal}{J. Phys. Chem. B} \textbf{\bibinfo{volume}{104}},
  \bibinfo{pages}{5430} (\bibinfo{year}{2000}).

\bibitem[{\citenamefont{Rousset et~al.}(1995)\citenamefont{Rousset, Aires,
  Renouprez, M\'elinon, P\'erez, Pellarin, Vialle, and Broyer}}]{rousset2000}
\bibinfo{author}{\bibfnamefont{J.~L.} \bibnamefont{Rousset}},
  \bibinfo{author}{\bibfnamefont{F.~J. C.~S.} \bibnamefont{Aires}},
  \bibinfo{author}{\bibfnamefont{A.}~\bibnamefont{Renouprez}},
  \bibinfo{author}{\bibfnamefont{P.}~\bibnamefont{M\'elinon}},
  \bibinfo{author}{\bibfnamefont{A.}~\bibnamefont{P\'erez}},
  \bibinfo{author}{\bibfnamefont{M.}~\bibnamefont{Pellarin}},
  \bibinfo{author}{\bibfnamefont{J.~L.} \bibnamefont{Vialle}},
  \bibnamefont{and} \bibinfo{author}{\bibfnamefont{M.}~\bibnamefont{Broyer}},
  \bibinfo{journal}{J. Chem. Phys.} \textbf{\bibinfo{volume}{102}},
  \bibinfo{pages}{8574} (\bibinfo{year}{1995}).

\bibitem[{\citenamefont{Gaudry et~al.}(2001)\citenamefont{Gaudry, Lerme,
  Cottancin, Pellarin, Pr\'evel, Treilleux, M\'elinon, and ans
  M.~Broyer}}]{gaudry2001}
\bibinfo{author}{\bibfnamefont{M.}~\bibnamefont{Gaudry}},
  \bibinfo{author}{\bibfnamefont{J.}~\bibnamefont{Lerme}},
  \bibinfo{author}{\bibfnamefont{E.}~\bibnamefont{Cottancin}},
  \bibinfo{author}{\bibfnamefont{M.}~\bibnamefont{Pellarin}},
  \bibinfo{author}{\bibfnamefont{B.}~\bibnamefont{Pr\'evel}},
  \bibinfo{author}{\bibfnamefont{M.}~\bibnamefont{Treilleux}},
  \bibinfo{author}{\bibnamefont{M\'elinon}}, \bibnamefont{and}
  \bibinfo{author}{\bibfnamefont{J.~L.~R.} \bibnamefont{ans M.~Broyer}},
  \bibinfo{journal}{Eur. Phys. J. D} \textbf{\bibinfo{volume}{16}},
  \bibinfo{pages}{201} (\bibinfo{year}{2001}).

\bibitem[{\citenamefont{Favre et~al.}(2004)\citenamefont{Favre, Stanescu,
  Dupuis, Bernstein, Epicier, M\'elinon, and P\'erez}}]{favre2004}
\bibinfo{author}{\bibfnamefont{L.}~\bibnamefont{Favre}},
  \bibinfo{author}{\bibfnamefont{S.}~\bibnamefont{Stanescu}},
  \bibinfo{author}{\bibfnamefont{V.}~\bibnamefont{Dupuis}},
  \bibinfo{author}{\bibfnamefont{E.}~\bibnamefont{Bernstein}},
  \bibinfo{author}{\bibfnamefont{T.}~\bibnamefont{Epicier}},
  \bibinfo{author}{\bibfnamefont{P.}~\bibnamefont{M\'elinon}},
  \bibnamefont{and} \bibinfo{author}{\bibfnamefont{A.}~\bibnamefont{P\'erez}},
  \bibinfo{journal}{Appl. Surf. Sci.} \textbf{\bibinfo{volume}{226}},
  \bibinfo{pages}{265} (\bibinfo{year}{2004}).

\bibitem[{\citenamefont{Favre et~al.}()\citenamefont{Favre, Dupuis, Bernstein,
  Stanescu, M\'elinon, P\'erez, Epicier, Simon, Tonnerre, and
  Babonneau}}]{favre2006}
\bibinfo{author}{\bibfnamefont{L.}~\bibnamefont{Favre}},
  \bibinfo{author}{\bibfnamefont{V.}~\bibnamefont{Dupuis}},
  \bibinfo{author}{\bibfnamefont{E.}~\bibnamefont{Bernstein}},
  \bibinfo{author}{\bibfnamefont{S.}~\bibnamefont{Stanescu}},
  \bibinfo{author}{\bibfnamefont{P.}~\bibnamefont{M\'elinon}},
  \bibinfo{author}{\bibfnamefont{A.}~\bibnamefont{P\'erez}},
  \bibinfo{author}{\bibfnamefont{T.}~\bibnamefont{Epicier}},
  \bibinfo{author}{\bibfnamefont{J.~P.} \bibnamefont{Simon}},
  \bibinfo{author}{\bibfnamefont{J.~M.} \bibnamefont{Tonnerre}},
  \bibnamefont{and}
  \bibinfo{author}{\bibfnamefont{D.}~\bibnamefont{Babonneau}},
  \bibinfo{howpublished}{unpublished}.

\bibitem[{\citenamefont{van Hardeveld and Hartog}(1969)}]{vanhardeveld1969}
\bibinfo{author}{\bibfnamefont{R.}~\bibnamefont{van Hardeveld}}
  \bibnamefont{and} \bibinfo{author}{\bibfnamefont{F.}~\bibnamefont{Hartog}},
  \bibinfo{journal}{Sur. Sci.} \textbf{\bibinfo{volume}{15}},
  \bibinfo{pages}{189} (\bibinfo{year}{1969}).

\bibitem[{\citenamefont{Jamet et~al.}(2000)\citenamefont{Jamet, Dupuis,
  M\'elinon, Guiraud, P\'erez, Wernsdorfer, Traverse, and
  Baguenard}}]{jamet2000}
\bibinfo{author}{\bibfnamefont{M.}~\bibnamefont{Jamet}},
  \bibinfo{author}{\bibfnamefont{V.}~\bibnamefont{Dupuis}},
  \bibinfo{author}{\bibfnamefont{P.}~\bibnamefont{M\'elinon}},
  \bibinfo{author}{\bibfnamefont{G.}~\bibnamefont{Guiraud}},
  \bibinfo{author}{\bibfnamefont{A.}~\bibnamefont{P\'erez}},
  \bibinfo{author}{\bibfnamefont{W.}~\bibnamefont{Wernsdorfer}},
  \bibinfo{author}{\bibfnamefont{A.}~\bibnamefont{Traverse}}, \bibnamefont{and}
  \bibinfo{author}{\bibfnamefont{B.}~\bibnamefont{Baguenard}},
  \bibinfo{journal}{Phys. Rev. B} \textbf{\bibinfo{volume}{62}},
  \bibinfo{pages}{493} (\bibinfo{year}{2000}).

\bibitem[{\citenamefont{Grolier et~al.}(1993)\citenamefont{Grolier, Renard,
  Bartenlian, Beauvillain, Chappert, Dupas, Ferré, Galtier, Kolb, Mulloy
  et~al.}}]{grolier1993}
\bibinfo{author}{\bibfnamefont{V.}~\bibnamefont{Grolier}},
  \bibinfo{author}{\bibfnamefont{D.}~\bibnamefont{Renard}},
  \bibinfo{author}{\bibfnamefont{B.}~\bibnamefont{Bartenlian}},
  \bibinfo{author}{\bibfnamefont{P.}~\bibnamefont{Beauvillain}},
  \bibinfo{author}{\bibfnamefont{C.}~\bibnamefont{Chappert}},
  \bibinfo{author}{\bibfnamefont{C.}~\bibnamefont{Dupas}},
  \bibinfo{author}{\bibfnamefont{J.}~\bibnamefont{Ferré}},
  \bibinfo{author}{\bibfnamefont{M.}~\bibnamefont{Galtier}},
  \bibinfo{author}{\bibfnamefont{E.}~\bibnamefont{Kolb}},
  \bibinfo{author}{\bibfnamefont{M.}~\bibnamefont{Mulloy}},
  \bibnamefont{et~al.}, \bibinfo{journal}{Phys. Rev. Lett.}
  \textbf{\bibinfo{volume}{71}}, \bibinfo{pages}{3023} (\bibinfo{year}{1993}).

\bibitem[{\citenamefont{Bruno and Chappert}(1991)}]{bruno1991}
\bibinfo{author}{\bibfnamefont{P.}~\bibnamefont{Bruno}} \bibnamefont{and}
  \bibinfo{author}{\bibfnamefont{C.}~\bibnamefont{Chappert}},
  \bibinfo{journal}{Phys. Rev. Lett.} \textbf{\bibinfo{volume}{67}},
  \bibinfo{pages}{1602} (\bibinfo{year}{1991}).

\bibitem[{\citenamefont{Allia et~al.}(2001)\citenamefont{Allia, Coisson,
  Tiberto, Vinai, Knobel, Novak, and Nunes}}]{allia2001}
\bibinfo{author}{\bibfnamefont{P.}~\bibnamefont{Allia}},
  \bibinfo{author}{\bibfnamefont{M.}~\bibnamefont{Coisson}},
  \bibinfo{author}{\bibfnamefont{P.}~\bibnamefont{Tiberto}},
  \bibinfo{author}{\bibfnamefont{F.}~\bibnamefont{Vinai}},
  \bibinfo{author}{\bibfnamefont{M.}~\bibnamefont{Knobel}},
  \bibinfo{author}{\bibfnamefont{M.~A.} \bibnamefont{Novak}}, \bibnamefont{and}
  \bibinfo{author}{\bibfnamefont{W.~C.} \bibnamefont{Nunes}},
  \bibinfo{journal}{Phys. Rev. B} \textbf{\bibinfo{volume}{64}},
  \bibinfo{pages}{144420} (\bibinfo{year}{2001}).

\bibitem[{\citenamefont{Kootte et~al.}(1991)\citenamefont{Kootte, Haas, and
  de~Groot}}]{kootte1991}
\bibinfo{author}{\bibfnamefont{A.}~\bibnamefont{Kootte}},
  \bibinfo{author}{\bibfnamefont{C.}~\bibnamefont{Haas}}, \bibnamefont{and}
  \bibinfo{author}{\bibfnamefont{R.~A.} \bibnamefont{de~Groot}},
  \bibinfo{journal}{J. Phys.: Condens. matter} \textbf{\bibinfo{volume}{3}},
  \bibinfo{pages}{1133} (\bibinfo{year}{1991}).

\bibitem[{\citenamefont{Fruchart et~al.}(2002)\citenamefont{Fruchart, Jubert,
  Meyer, Klaua, Barthel, and Kirschner}}]{fruch2002}
\bibinfo{author}{\bibfnamefont{O.}~\bibnamefont{Fruchart}},
  \bibinfo{author}{\bibfnamefont{P.~O.} \bibnamefont{Jubert}},
  \bibinfo{author}{\bibfnamefont{C.}~\bibnamefont{Meyer}},
  \bibinfo{author}{\bibfnamefont{M.}~\bibnamefont{Klaua}},
  \bibinfo{author}{\bibfnamefont{J.}~\bibnamefont{Barthel}}, \bibnamefont{and}
  \bibinfo{author}{\bibfnamefont{J.}~\bibnamefont{Kirschner}},
  \bibinfo{journal}{J. Mag. Mag. Mat.} \textbf{\bibinfo{volume}{239}},
  \bibinfo{pages}{224} (\bibinfo{year}{2002}).

\bibitem[{\citenamefont{Grange et~al.}(2000)\citenamefont{Grange, Galanakis,
  Alouani, Maret, Kappler, and Rogalev}}]{grange2000}
\bibinfo{author}{\bibfnamefont{W.}~\bibnamefont{Grange}},
  \bibinfo{author}{\bibfnamefont{I.}~\bibnamefont{Galanakis}},
  \bibinfo{author}{\bibfnamefont{M.}~\bibnamefont{Alouani}},
  \bibinfo{author}{\bibfnamefont{M.}~\bibnamefont{Maret}},
  \bibinfo{author}{\bibfnamefont{J.~P.} \bibnamefont{Kappler}},
  \bibnamefont{and} \bibinfo{author}{\bibfnamefont{A.}~\bibnamefont{Rogalev}},
  \bibinfo{journal}{Phys. Rev. B} \textbf{\bibinfo{volume}{62}},
  \bibinfo{pages}{1157} (\bibinfo{year}{2000}).

\bibitem[{\citenamefont{Dupuis et~al.}(2003)\citenamefont{Dupuis, Jamet, Favre,
  Tuaillon-Combes, M\'elinon, and P\'erez}}]{dupuis2003}
\bibinfo{author}{\bibfnamefont{V.}~\bibnamefont{Dupuis}},
  \bibinfo{author}{\bibfnamefont{M.}~\bibnamefont{Jamet}},
  \bibinfo{author}{\bibfnamefont{L.}~\bibnamefont{Favre}},
  \bibinfo{author}{\bibfnamefont{J.}~\bibnamefont{Tuaillon-Combes}},
  \bibinfo{author}{\bibfnamefont{P.}~\bibnamefont{M\'elinon}},
  \bibnamefont{and} \bibinfo{author}{\bibfnamefont{A.}~\bibnamefont{P\'erez}},
  \bibinfo{journal}{J. Vac. Sci. Technol. A} \textbf{\bibinfo{volume}{21}},
  \bibinfo{pages}{1519} (\bibinfo{year}{2003}).

\bibitem[{\citenamefont{Wernsdorfer et~al.}(1997)\citenamefont{Wernsdorfer,
  Orozco, Hasselbach, Benoit, Barbara, Demoncy, Loiseau, Pascard, and
  Mailly}}]{wernsdorfer1997}
\bibinfo{author}{\bibfnamefont{W.}~\bibnamefont{Wernsdorfer}},
  \bibinfo{author}{\bibfnamefont{E.~B.} \bibnamefont{Orozco}},
  \bibinfo{author}{\bibfnamefont{K.}~\bibnamefont{Hasselbach}},
  \bibinfo{author}{\bibfnamefont{A.}~\bibnamefont{Benoit}},
  \bibinfo{author}{\bibfnamefont{B.}~\bibnamefont{Barbara}},
  \bibinfo{author}{\bibfnamefont{N.}~\bibnamefont{Demoncy}},
  \bibinfo{author}{\bibfnamefont{A.}~\bibnamefont{Loiseau}},
  \bibinfo{author}{\bibfnamefont{H.}~\bibnamefont{Pascard}}, \bibnamefont{and}
  \bibinfo{author}{\bibfnamefont{D.}~\bibnamefont{Mailly}},
  \bibinfo{journal}{Phys. Rev. Lett.} \textbf{\bibinfo{volume}{78}},
  \bibinfo{pages}{1791} (\bibinfo{year}{1997}).

\bibitem[{\citenamefont{Stoner and Wohlfarth}(1948)}]{stoner1948}
\bibinfo{author}{\bibfnamefont{E.~C.} \bibnamefont{Stoner}} \bibnamefont{and}
  \bibinfo{author}{\bibfnamefont{E.~P.} \bibnamefont{Wohlfarth}},
  \bibinfo{journal}{Phil. Trans. Roy. Soc. London} \textbf{\bibinfo{volume}{240
  A}}, \bibinfo{pages}{599} (\bibinfo{year}{1948}).

\bibitem[{\citenamefont{Chantrell et~al.}(1994)\citenamefont{Chantrell,
  Lyberatos, El-Hilo, and O'Grady}}]{chantrell1994}
\bibinfo{author}{\bibfnamefont{R.~W.} \bibnamefont{Chantrell}},
  \bibinfo{author}{\bibfnamefont{A.}~\bibnamefont{Lyberatos}},
  \bibinfo{author}{\bibfnamefont{M.}~\bibnamefont{El-Hilo}}, \bibnamefont{and}
  \bibinfo{author}{\bibfnamefont{K.}~\bibnamefont{O'Grady}},
  \bibinfo{journal}{J. Appl. Phys.} \textbf{\bibinfo{volume}{76}},
  \bibinfo{pages}{6407} (\bibinfo{year}{1994}).

\bibitem[{\citenamefont{Street and Brown}(1994)}]{street1994}
\bibinfo{author}{\bibfnamefont{R.}~\bibnamefont{Street}} \bibnamefont{and}
  \bibinfo{author}{\bibfnamefont{S.~D.} \bibnamefont{Brown}},
  \bibinfo{journal}{J. Appl. Phys.} \textbf{\bibinfo{volume}{76}},
  \bibinfo{pages}{6386} (\bibinfo{year}{1994}).

\bibitem[{\citenamefont{Andersson et~al.}(1997)\citenamefont{Andersson,
  Djurberg, Jonsson, Svedlindh, and Nordblad}}]{andersson1997}
\bibinfo{author}{\bibfnamefont{J.~O.} \bibnamefont{Andersson}},
  \bibinfo{author}{\bibfnamefont{C.}~\bibnamefont{Djurberg}},
  \bibinfo{author}{\bibfnamefont{T.}~\bibnamefont{Jonsson}},
  \bibinfo{author}{\bibfnamefont{P.}~\bibnamefont{Svedlindh}},
  \bibnamefont{and} \bibinfo{author}{\bibfnamefont{P.}~\bibnamefont{Nordblad}},
  \bibinfo{journal}{Phys. Rev. B} \textbf{\bibinfo{volume}{56}},
  \bibinfo{pages}{13983} (\bibinfo{year}{1997}).

\bibitem[{\citenamefont{Hillebrands and Dutcher}(1993)}]{hillebrands1993}
\bibinfo{author}{\bibfnamefont{B.}~\bibnamefont{Hillebrands}} \bibnamefont{and}
  \bibinfo{author}{\bibfnamefont{J.~R.} \bibnamefont{Dutcher}},
  \bibinfo{journal}{Phys. Rev. B} \textbf{\bibinfo{volume}{47}},
  \bibinfo{pages}{6126} (\bibinfo{year}{1993}).

\bibitem[{\citenamefont{Morel et~al.}(2004)\citenamefont{Morel, Brenac, and
  Portemont}}]{morel2004}
\bibinfo{author}{\bibfnamefont{R.}~\bibnamefont{Morel}},
  \bibinfo{author}{\bibfnamefont{A.}~\bibnamefont{Brenac}}, \bibnamefont{and}
  \bibinfo{author}{\bibfnamefont{C.}~\bibnamefont{Portemont}},
  \bibinfo{journal}{J. Appl. Phys.} \textbf{\bibinfo{volume}{95}},
  \bibinfo{pages}{3757} (\bibinfo{year}{2004}).

\bibitem[{\citenamefont{Dobrynin et~al.}(2005)\citenamefont{Dobrynin, Ievlev,
  Temst, Lievens, Margueritat, Gonzalo, Afonso, Zhou, Vantomme, Piscopiello
  et~al.}}]{dobrynin2005}
\bibinfo{author}{\bibfnamefont{A.~N.} \bibnamefont{Dobrynin}},
  \bibinfo{author}{\bibfnamefont{D.~N.} \bibnamefont{Ievlev}},
  \bibinfo{author}{\bibfnamefont{K.}~\bibnamefont{Temst}},
  \bibinfo{author}{\bibfnamefont{P.}~\bibnamefont{Lievens}},
  \bibinfo{author}{\bibfnamefont{J.}~\bibnamefont{Margueritat}},
  \bibinfo{author}{\bibfnamefont{J.}~\bibnamefont{Gonzalo}},
  \bibinfo{author}{\bibfnamefont{C.~N.} \bibnamefont{Afonso}},
  \bibinfo{author}{\bibfnamefont{S.~Q.} \bibnamefont{Zhou}},
  \bibinfo{author}{\bibfnamefont{A.}~\bibnamefont{Vantomme}},
  \bibinfo{author}{\bibfnamefont{E.}~\bibnamefont{Piscopiello}},
  \bibnamefont{et~al.}, \bibinfo{journal}{Appl. Phys. Lett.}
  \textbf{\bibinfo{volume}{87}}, \bibinfo{pages}{12501} (\bibinfo{year}{2005}).

\bibitem[{\citenamefont{Hong et~al.}(2006)\citenamefont{Hong, Leo, Smith, and
  Berkowitz}}]{hong2006}
\bibinfo{author}{\bibfnamefont{J.~I.} \bibnamefont{Hong}},
  \bibinfo{author}{\bibfnamefont{T.}~\bibnamefont{Leo}},
  \bibinfo{author}{\bibfnamefont{D.~J.} \bibnamefont{Smith}}, \bibnamefont{and}
  \bibinfo{author}{\bibfnamefont{A.~E.} \bibnamefont{Berkowitz}},
  \bibinfo{journal}{Phys. Rev. Lett.} \textbf{\bibinfo{volume}{96}},
  \bibinfo{pages}{117204} (\bibinfo{year}{2006}).

\end{thebibliography}

\end{document}